\documentclass[conference, a4paper]{IEEEtran}
\usepackage[utf8]{inputenc}
\usepackage{mathtools} 
\usepackage{blindtext, graphicx}
\usepackage{xcolor}
\usepackage{diagbox} 
\usepackage{multirow}

\newcommand{\NCP}{N_\mathrm{CP}}

\newcommand{\tth}{$^{\mathrm{th}}$}

\begin{document}
\title{Analytical study of 5G NR eMBB co-existence}
\author{\IEEEauthorblockN{David Demmer\IEEEauthorrefmark{1}\IEEEauthorrefmark{2}, Robin Gerzaguet\IEEEauthorrefmark{3}, Jean-Baptiste Dor\'e\IEEEauthorrefmark{1}, Didier Le Ruyet\IEEEauthorrefmark{2} }
\IEEEauthorblockA{\IEEEauthorrefmark{1}CEA-Leti, Minatec Campus, Grenoble, France \\ \{david.demmer, jean-baptiste.dore\}@cea.fr}
\IEEEauthorblockA{\IEEEauthorrefmark{2}Conservatoire National des Arts et M\'etiers, Paris, France, didier.le\_ruyet@cnam.fr}
\IEEEauthorblockA{\IEEEauthorrefmark{3}Univ Rennes, CNRS, IRISA, France, robin.gerzaguet@irisa.fr}\\
	}
\maketitle

\begin{abstract}
3GPP release 15 focusing on 5G general outline has been published in December 2017. The major difference with respect to currently deployed LTE is the support of various physical layer numerologies. Making the physical layer scalable allows to properly address new services such as low latency  or  millimeter communications. However it poses the problem of numerology coexistence. Indeed the orthogonality of the OFDM waveform is broken by the use of different subcarrier spacings and therefore multiplexed communications may interfere with each others. A first and simple solution to limit the distortion is to consider guard bands. In this paper, the authors develop analytical metrics to quantify the level of distortion induced by 5G multi-service multiplexing. Besides, general comments and guard band dimensioning are carried out.  It is shown that high interference rejection needs to be associated with side lobe reduction techniques to favor an efficient bandwidth use.
\end{abstract}


\section{Introduction}
Cellular networks have evolved a lot since their creation.  They were initially designed to support low-data rate Human-to-Human communications. Nowadays 4G LTE also provides broadband Internet access with a huge data consumption on the downlink. Tomorrow, the cellular networks are expected to support Machine-Type-Communications (MTC) as well. It brings a set of new typical applications to be supported by the network from sporadic communications with a long idle-state to critical communications with low latency constraints. The currently deployed 4G LTE shows intrinsic limits to support the heterogeneous services \cite{NGMN}\cite{5GPPP}. It thus pushes the forthcoming wireless technologies to favor superior flexibility which paves the way to 5G New Radio (NR).

Release 15 of the 3GPP has been published in December 2017 and is mainly dedicated to 5G NR Enhanced Mobile Broadband (eMBB) and Fixed Wireless Access (FWA)\cite{3gpp.38.300}. It defines the complete framework for the physical layer and exhibits structural differences with legacy LTE. First, it addresses higher frequencies with targeted carrier frequencies up to $52.6$ GHz. Then, the typical bandwidth is increased from $20$ MHz  to $400$ MHz. Last but not least, the PHY layer is designed to support various configurations. Multiplexing different numerologies allows the service to choose between a set of supported subcarrier spacing (SCS) and symbol durations that better fit its needs. However  the problem of service coexistence arise. Indeed, multiplexing various SCS/symbol durations breaks the OFDM orthogonality and thus is source of interference.  One approach to limit the interference is the use of large guard bands between distinct-numerology transmissions which however lowers the bandwidth use. This practice is not compatible with the initial vision of 5G which aims at maximizing the bandwidth use to properly support the huge expected traffic\cite{5GPPP}. 

The main contribution of this paper is the examination of the 5G NR numerology  coexistence issue. The level of induced distortion is analytically expressed which leads to an analysis and an evaluation of the guard band size required to satisfy given targets quality of signals. 


The remainder of the paper is structured as follows. Section II is dedicated to the system model. The 5G NR numerologies are introduced and the OFDM transmissions techniques as well. Section III is dedicated to the derivation of the distortion level indicator. The Mean Square Error (MSE) has been chosen as indicator. Then, the analysis and guard band size evaluation is conducted in Section IV. Section V gives the conclusion and perspectives.

\section{System Model}

\subsection{5G NR standard and numerologies}

\begin{table}[bt]
	\centering
	\caption{5G NR numerology.}
	\label{tab:5GNR-Data-Config}
	\begin{tabular}{|c|c|c|c|}
		\hline
		$\mu$ & SCS [kHz] & RB band [kHz] & Symbol + CP length [samples] \\ \hline
		0     & 15.0       & 180.0              &      4096 + 288            \\ \hline
		1     & 30.0       & 360.0              &      2048 + 144            \\ \hline
		2     & 60.0       & 720.0             &  1024 + 72             \\ \hline
		3     & 120.0      & 1440.0              & 512 + 36                \\ \hline
		4     & 240.0      & 2880.0            &  256 + 18               \\ \hline
	\end{tabular}
\end{table}

Regarding physical layer specifications, the major innovation of release $15$ with respect to former standards is the support of various numerologies \cite{3gpp.38.300} which allows scalable SCS and symbol duration. By doing so, challenges of 5G can be properly addressed. Indeed shortening symbols reduces the latency required for some Ultra Reliable and Low Latency Communications\cite{Lon16}. Moreover, enlarging the SCS increases the robustness against the Doppler effect occurring in mobility scenarios and strong phase noise of millimeter wave frequencies \cite{Zaidi2016a}. 

Therefore, five different numerologies will be supported as shown in Table \ref{tab:5GNR-Data-Config}. They are identified with a numerology index $\mu$ which leads to a SCS equal to $15\times 2^{\mu}$ Hz. Numerologies \mbox{$\mu= \{0,1,2\}$} target sub-6 GHz communications and $\mu= \{2,3,4\}$ are rather dedicated to higher frequencies.
It seems worth noticing that a Resource Block (RB) is still defined as a group of $12$ contiguous subcarriers. Hence the RB bandwidth in release 15 depends on the numerology. Besides, the sampling frequency is set to $61.44$ MHz for all numerologies. 

\begin{figure}[tb]
	\centering
	\includegraphics[width=1\columnwidth, keepaspectratio]{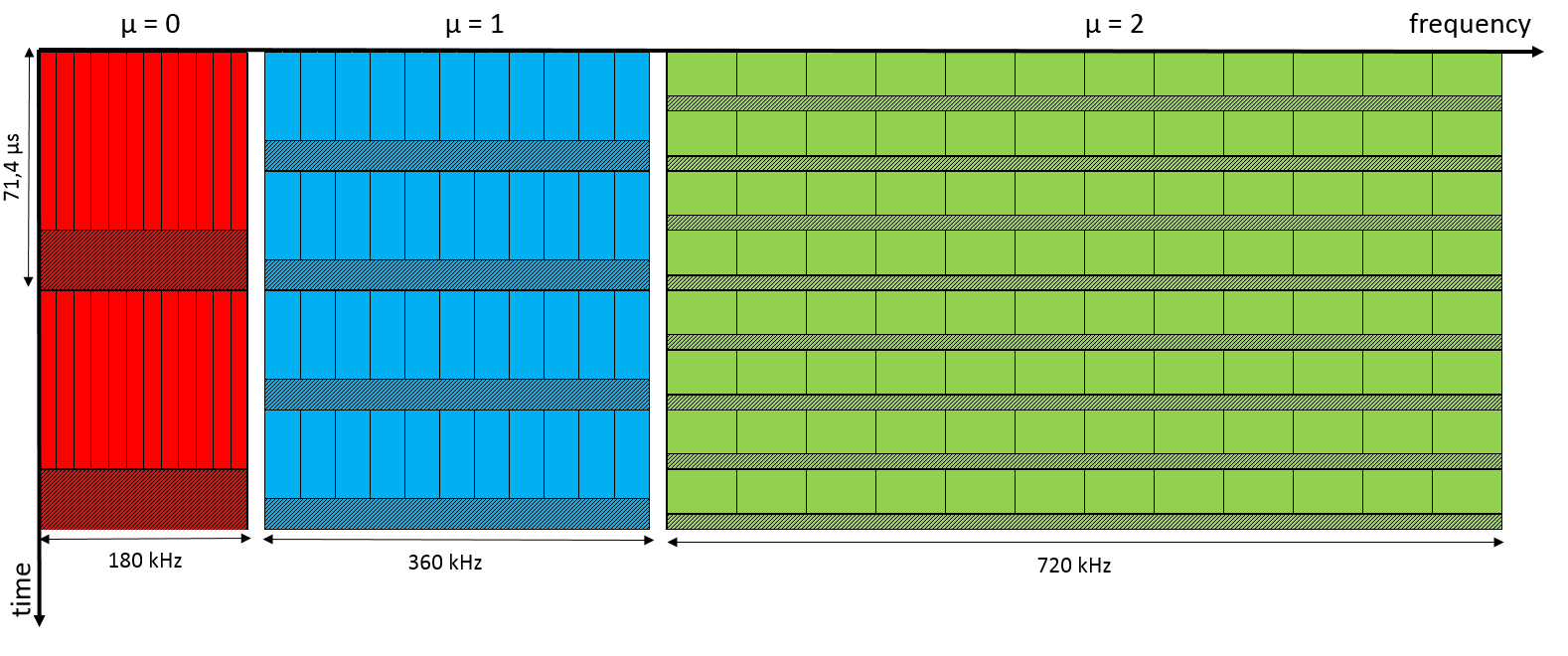}
	\caption{Time frequency illustration of the 5G NR frame structure with different numerologies }
	\label{fig:tf}
\end{figure}

Symbol alignment in time can still be ensured as the symbol duration linearly decreases with $\mu$ as depicted in \mbox{Figure \ref{fig:tf}}. Resource allocation in time can therefore be performed according to a frame system as in LTE. Downlink and uplink communications are organized into frames of $10$ ms \cite{3gpp.38.300}.

The total cell bandwidth is split into Bandwidth Parts (BWP). Bandwidth adaptation is performed by configuring the user equipment (UE) with the BWPs. By doing so, the cell bandwidth can be larger than the maximum UE achievable bandwidth \cite{3gpp.38.300}. Besides, the different services can be supported in the same cell bandwidth. Indeed, even if each BWP supports an unique numerology\footnote{multiple-numerology BWP is not addressed in current releases}, adjacent BWPs can address distinct services. 

\subsection{Notations}
From now and for the rest of the paper, we propose the following notations to respectively express the useful symbol $N^\mu$, the cyclic prefix (CP) $\NCP^\mu $ and entire symbol lengths $N_e^\mu$ as function of $\mu$.
\begin{align}
N^\mu &=N2^{-\mu} = 4096 \times 2^{-\mu} \\
\NCP^\mu &=\NCP2^{-\mu} =  288 \times 2^{-\mu}\\
N_e^\mu &= N^\mu  + \NCP^\mu 
\end{align}

\subsection{Transmitter scheme}
The conventional CP-OFDM transmitted signal can be expressed as in (\ref{eq:cpofdm-tx}) where $a_{f,n}$ denote the transmitted constellation symbol over the $f$\tth subcarrier at time instant $n$, $\Omega$ the set of the active subcarriers, $T = (N^\mu + \NCP^\mu)T_s$ the length of one complete symbol where $T_s$ denotes the sampling period, $T_{\mathrm{CP}} ^ \mu= \NCP^\mu T_s$ the CP length, $\Delta_f^\mu = (T_s N^\mu)^{-1}$ the subcarrier spacing for numerology $\mu$ and  $\Pi_T[t]$ the gate function of length $T$ centered in $T/2$. 
\begin{equation}
s_\mu(t) =  \frac{1}{\sqrt{N^\mu}}\sum_{n}\sum_{f \in \Omega } a_{f,n}^\mu \Pi_{T}(t-nT) e^{j 2 \pi(t-T_{\mathrm{CP}}^\mu)f\Delta_f^\mu}
\label{eq:cpofdm-tx}
\end{equation}

\subsection{Receiver scheme}
The incoming signal is first discretized with a sampling period $T_s$. By setting $s_\mu\left [ l \right ]  = s_\mu(l T_s)$, the conventional CP-OFDM received symbols can be expressed as in (\ref{eq:cpofdm-rx}).
\begin{equation}
\hat{a}_{f,n}^\mu =  \frac{1}{\sqrt{N^\mu}}\sum_{k=0}^{N^\mu-1} s_\mu\left [nN_\mathrm{e}^ \mu + \NCP^\mu  + k\right ] e^{-j \frac{2 \pi}{N^\mu}kf}
\label{eq:cpofdm-rx}
\end{equation}

The expression (\ref{eq:cpofdm-rx}) is given for distortion-free and noiseless propagation. Indeed, as the main purpose of this study is to analyze the interference induced by 5G NR numerologies, other sources of distortion such as multi-path channels are omitted. The study of a realistic 5G transmission and the comparison with the results derived in this paper is left free for future studies. 

Besides, the paper only addresses CP-OFDM communications and therefore focuses on the downlink. Nonetheless, the authors strongly believe that the results can be easily extended and verified for DFT-s-OFDM and therefore the uplink as well. 

Last but not least, power control is not considered neither in the proposed study. There are already many transmission parameters that are taken into consideration and therefore the authors have preferred limiting their number and omitting power control impact. The paper somehow considers the worst case where the User Equipment (UE) is at equal distance of Base Stations (BS) delivering different services. It can be either the same BS addressing two distinct services (\text{i.e.} numerologies) over adjacent BWPs or an UE at the edge of cells of same size providing distinct services. 

\section{Distortion Analysis}
\begin{figure}[tb]
	\centering
	\includegraphics[width=0.95\columnwidth, keepaspectratio]{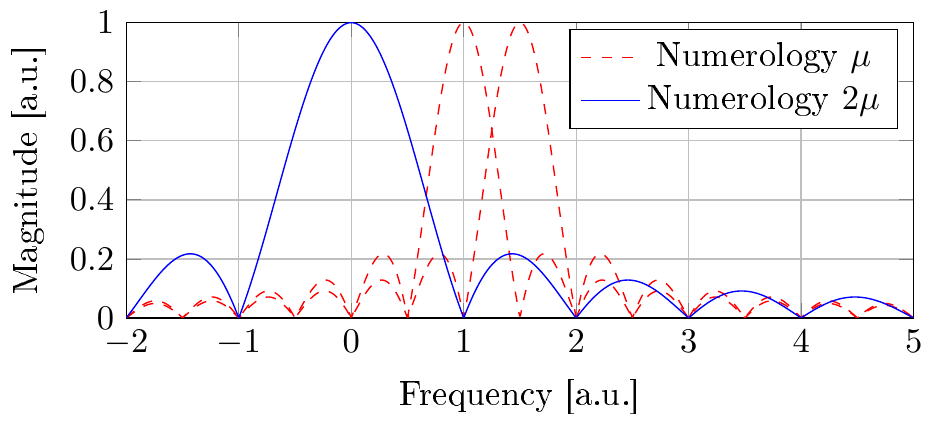}
	\caption{Sine cardinal of two distinct numerologies interfering  with each other (a.u. stands for arbitrary units) }
	\label{fig:sinc}
\end{figure}

In LTE, received signals remain orthogonal at the receiver sde from each other as long as they are synchronized within the CP duration and share the same numerology (subcarrier spacing, symbol length). According to 5G NR specifications, adjacent BWPs may address distinct services (different parameters $\mu$). Hence, they may interfere with each others even if proper frequency allocation and perfect synchronization at the receiver side are ensured. Indeed, frequency tones may not properly fit in zeros of sine cardinal (sinc) response of adjacent tones as illustrated in Figure \ref{fig:sinc}. Ensuring that frequency tones always corresponds to the zeros of the sinc response of all other tones would significantly compromise the bandwidth use efficiency and is therefore not acceptable.  

In this paper, we propose to express and quantify the level of distortion induced by the multiplexing of different services. The MSE has been chosen as distortion indicator. 

For the sake of the analysis, let us assume a single-tone (\textit{i.e.} on subcarrier) interferer working at frequency $f_i$ and with numerology ${\mu}_i$ and a single-tone user of interest working at frequency $f_u$ and with numerology ${\mu}_u$.  Extension to multiple allocated subcarriers will be provided later on.

The resulting MSE is defined in (\ref{eq:mse_def}) and developed in (\ref{eq:mse-dvlp}). 
\begin{equation}
\mathrm{MSE}^{{\mu}_i \rightarrow {\mu}_u} [\Delta_\mathrm{gb} ] =  \frac{ E_n \left [ \left | \hat{a}_ {f_u,n}^{{\mu}_u} - a_{f_u,n}^{{\mu}_u}\right |^2 \right ]}{\sigma_a^2}
\label{eq:mse_def}
\end{equation}

\begin{table*}[!bt]
	\begin{align}
	\mathrm{MSE}^{{\mu}_i \rightarrow {\mu}_u} [\Delta_\mathrm{gb} ] &= \frac{1}{N^{{\mu}_u}\sigma_a^2} E_n\left [\left| \sum_{l=0}^{N^{{\mu}_u}-1}s_{{\mu}_i}[l+\NCP^{{\mu}_u}+nN_e^{{\mu}_u}] e^{-j \frac{2 \pi}{N^{{\mu}_u}}lf_u} \right |^2  \right ]\nonumber \\
	& = \frac{1}{N^{{\mu}_u}N^{{\mu}_i}} \left [ \left|  \sum_{l=0}^{L-1} e^{j \frac{2\pi}{N}l \Delta_\mathrm{gb} }    \right |^2  + \sum_{k=0}^{\lceil Q \rceil-2}  \left| \sum_{l=L+kN_e^{{\mu}_i}}^{L+(k+1)N_e^{{\mu}_i}-1}  e^{j \frac{2\pi}{N}l \Delta_\mathrm{gb} }     \right |^2 \right ] \nonumber \\
	& = \frac{1}{N^{{\mu}_i}N^{{\mu}_u}} \left [ |D_L \left( \Delta_\mathrm{gb}  \right ) |^2+(\lceil Q \rceil-1) |D_{N_e^{{\mu}_i}} \left( \Delta_\mathrm{gb}  \right )  |^2\right]
	\label{eq:mse-dvlp}
	\end{align}
\end{table*}

Regarding the notations, \mbox{$\Delta_\mathrm{gb} =  \left  |f_i2^{{\mu}_i}-f_u2^{{\mu}_u}\right |$} is the guard band,  \mbox{$Q = 2^{{\mu}_i-{\mu}_u}$} is the number of OFDM symbols with numerology ${\mu}_i$ interfering with one symbol of numerology ${\mu}_u$, \mbox{$E_n\left [ . \right ]$} is the expectation operator applied in time, \mbox{$\lceil . \rceil$} is the ceiling operator, $L=N_e^{{\mu}_i}-\NCP^{{\mu}_u}$ and $D_L\left (f \right )$ are the partial sum of the Dirichlet Kernel expressed in $f$ defined by: 
\begin{equation}
D_L(f) = \sum_{l=0}^{L-1} e^{j \frac{2 \pi}{N}lf}
\label{eq:Dirichlet-K}
\end{equation}
The development of the expression (\ref{eq:mse-dvlp}) relies on the independence in time of the transmitted symbols and the separation of the $Q$ interfering OFDM symbols. 

The MSE expression (\ref{eq:mse-dvlp}) expresses the distortion level induced by a one-tone interferer over a unique-subcarrier band of interest with a guard band $\Delta_\mathrm{gb} $. One can observe than the distortion can be decomposed into $\lceil Q \rceil$ terms corresponding to the $\lceil Q \rceil$ interfering symbols overlapping with each symbol of interest. 

It is now possible to determine the level of distortion induced by $N_{\mathrm{int}}$ subcarriers over a sub-carrier spaced by a guard band $\Delta_\mathrm{gb} $, the contributions of each interfering sub-carrier can be summed (assuming that constellation symbols are independent). The resulting expression is given in (\ref{eq:mse_rb_def}). 

\begin{equation}
\mathrm{MSE}_{N_{\mathrm{int}}}^{{\mu}_i \rightarrow {\mu}_u} [\Delta_\mathrm{gb} ] = \sum_{k=0}^{\mathclap{N_{\mathrm{int}}-1}} \mathrm{MSE}^{{\mu}_i \rightarrow {\mu}_u} [\Delta_\mathrm{gb} +kQ]
\label{eq:mse_rb_def}
\end{equation}

It is also possible to determine the average level of distortion induced by $N_{\mathrm{int}}$ subcarriers over a full RB spaced by a guard band $\Delta_\mathrm{gb} $ as given in (\ref{eq:mse_rb_rb_def}). Even if this indicator can be interesting at the system level, it won't be used for this study. Indeed information is lost by averaging the level of distortion over multiple subcarriers which compromises the analysis. Generally speaking, the closest subcarrier to the interferer RBs suffers from the biggest distortion. Therefore, to the authors' point of view, considering only one useful subcarrier is more relevant. 
\begin{equation}
\mathrm{MSE}_{N_{\mathrm{int}} \rightarrow \mathrm{RB}}^{{\mu}_i \rightarrow {\mu}_u} [\Delta_\mathrm{gb} ] =  \frac{1}{12}\sum_{k=0}^{11} \mathrm{MSE}_{N_{\mathrm{int}}}^{{\mu}_i \rightarrow {\mu}_u} [\Delta_\mathrm{gb} + k2^{{\mu}_u}] 
\label{eq:mse_rb_rb_def}
\end{equation}

\section{Closed-form performance}
\subsection{Interference level}

\begin{figure*}[!ht]
	\centering
	\includegraphics[width=0.95\textwidth, keepaspectratio]{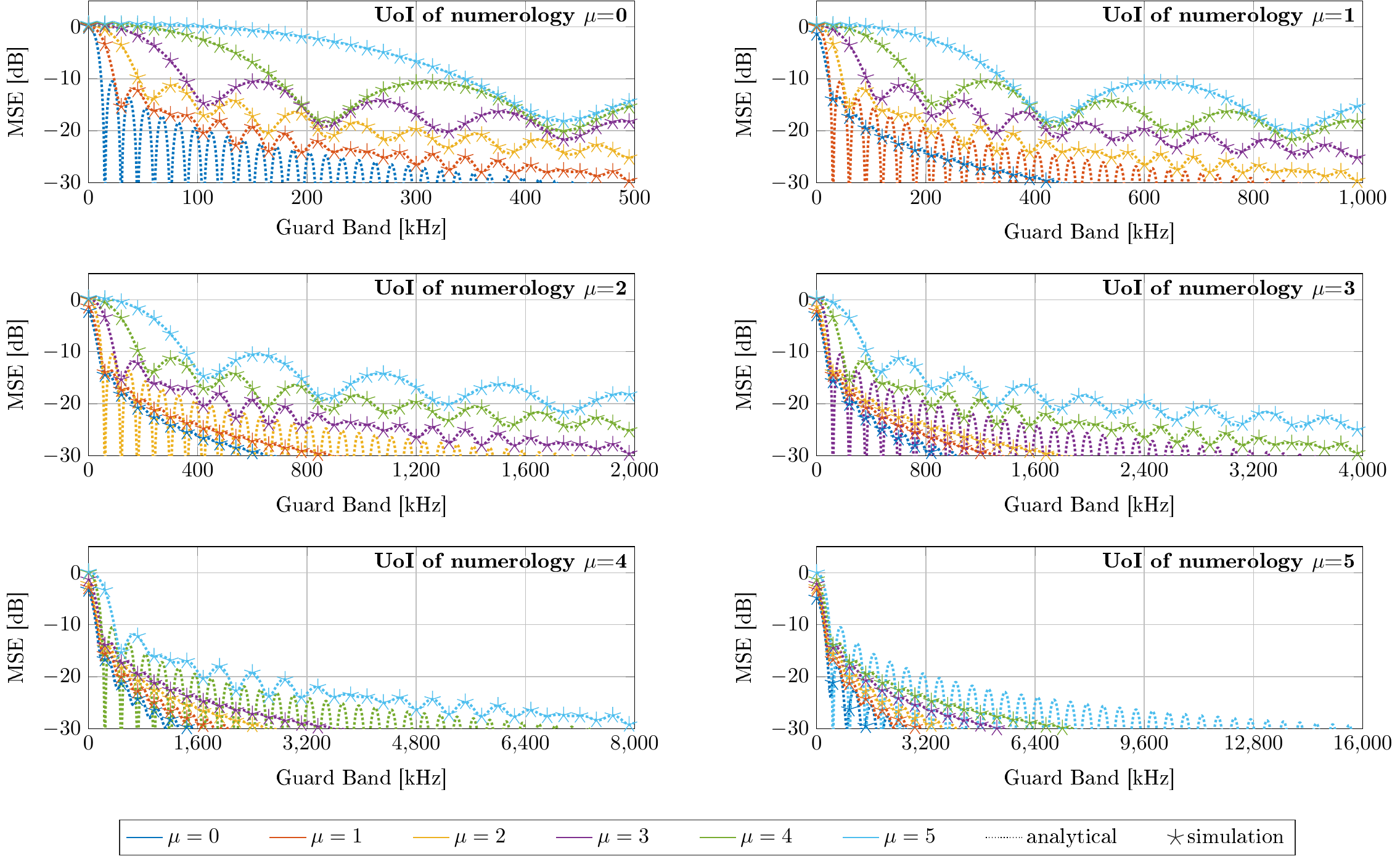}
	\caption{Level of distortion induced by 1-RB interferer over 1 sub-carrier user of interest according to guard band spacing the two users for $\mu=\{0,1,2,3,4,5\}$.}
	\label{fig:verif_ofdm}
\end{figure*}

The MSE expression developed in (\ref{eq:mse_rb_def}) is first verified and compared to simulation results for a whole RB ($N_{\mathrm{int}} =12$) as shown in Figure \ref{fig:verif_ofdm}. 

To begin with, one can observe that the analytical expression perfectly matches to the simulation results. Besides, when the interferer and the user of interest share the same numerology (whatever it is), it exists a guard band (and all its multiple) which guarantees the orthogonality between the two users. It is a well-known result used in LTE and OFDMA to multiplex several users \cite{OFDMA}. The minimum guard band that ensures the orthogonality is defined by the ratio between the sampling frequency and the total number of sub-carriers.

As expected, the result no longer holds when different numerologies are considered. They may even significantly interfere from each other even when large guard band spaces the two streams. It is worth pointing out that the distortion induced by lower numerologies ($Q <1$) impacts less than higher numerologies. Moreover, for higher numerologies, the bigger $Q$ is, the more distortion occurs. Indeed, $\lceil Q \rceil $ corresponds to the number of interfering waveform symbols, therefore it makes sense that the higher  $\lceil Q \rceil $ is, the more distortion is induced. 

It seems worth observing that it is possible to simplify the model by reducing the number of degrees of freedom by one. Indeed by looking at Figure \ref{fig:verif_ofdm}, one can observe that the approximation (\ref{eq:approx_mse}) is valid for any $\alpha$ such as \mbox{$(\mu_i+\alpha)\times(\mu_u+\alpha) \in [0,5]^2$}. It can be verified that the approximation still holds for $N_\mathrm{int} \neq 12$. It means that if the MSE is expressed for a normalized guard band (for instance the number of subcarrier instead of in frequential units), the MSE does not depend on the two numerology indexes $\mu_i$ and $\mu_u$ but only on their ratio $Q$. By considering this approximation, the MSE is function of $N_\mathrm{int}$ the number of interferer subcarriers, $Q$ the numerology index ratio and the guard band expressed in number of subcarriers. 
\begin{equation}
\mathrm{MSE}_{12}^{{\mu}_i +\alpha \rightarrow {\mu}_u + \alpha} [2^\alpha\Delta_\mathrm{gb} ] \approx \mathrm{MSE}_{12}^{{\mu}_i \rightarrow {\mu}_u} [\Delta_\mathrm{gb} ]  
\label{eq:approx_mse}
\end{equation}

\begin{figure*}[tb]
	\centering
	\includegraphics[width=0.95\columnwidth, keepaspectratio]{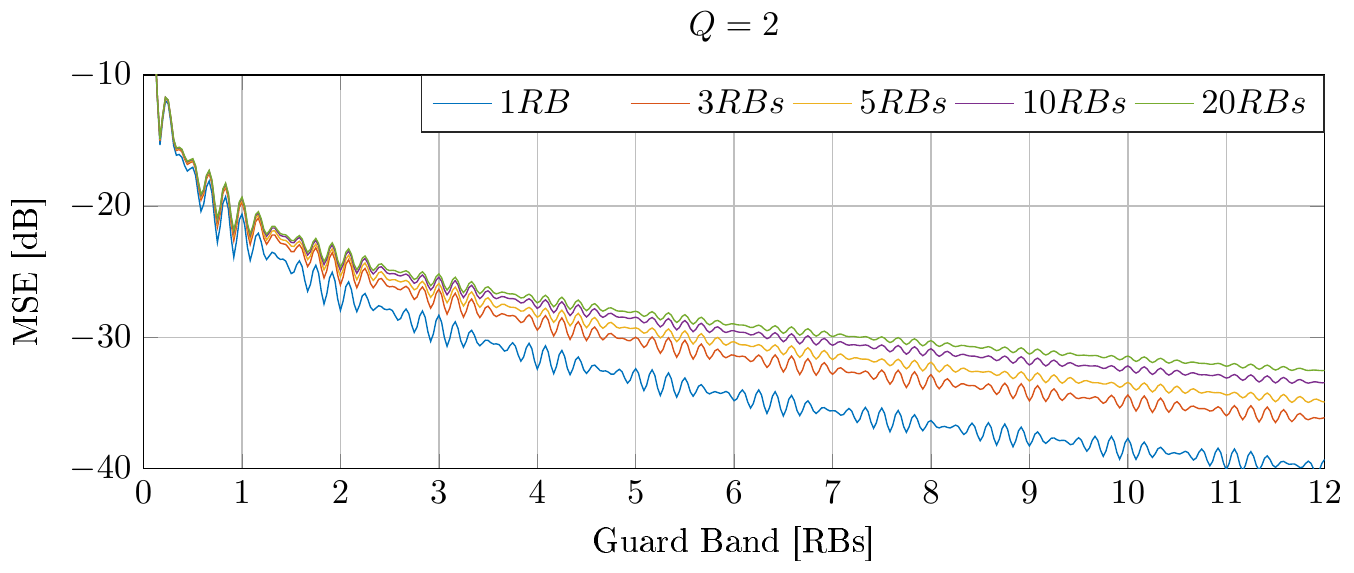}
	\includegraphics[width=0.95\columnwidth, keepaspectratio]{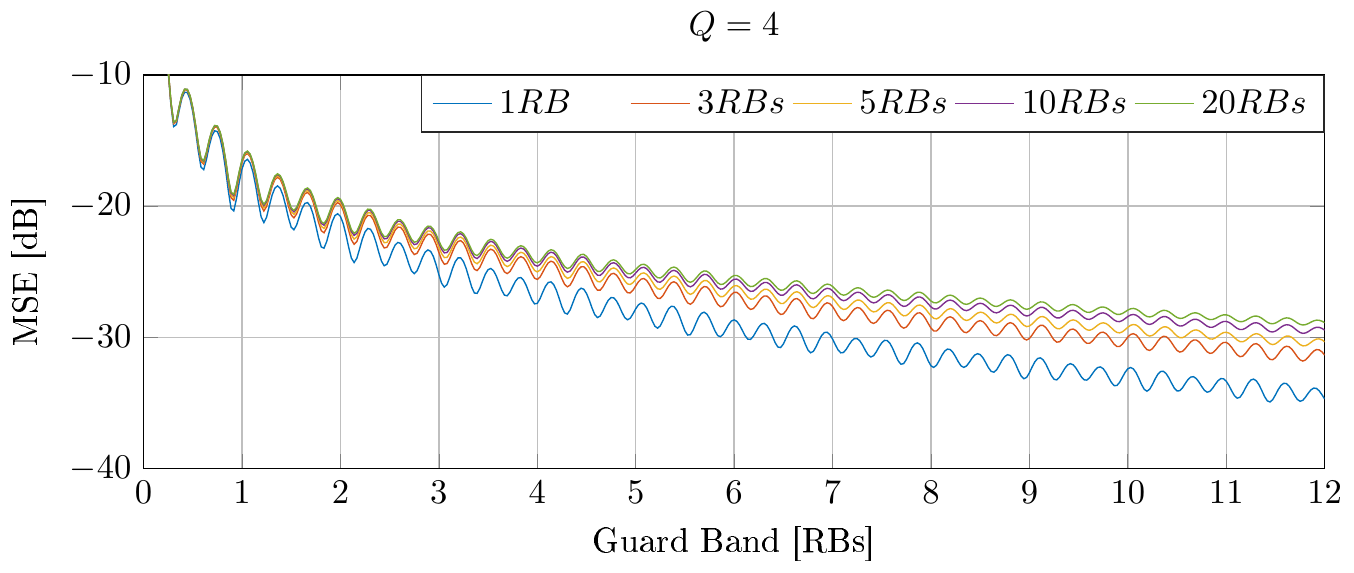}
	\includegraphics[width=0.95\columnwidth, keepaspectratio]{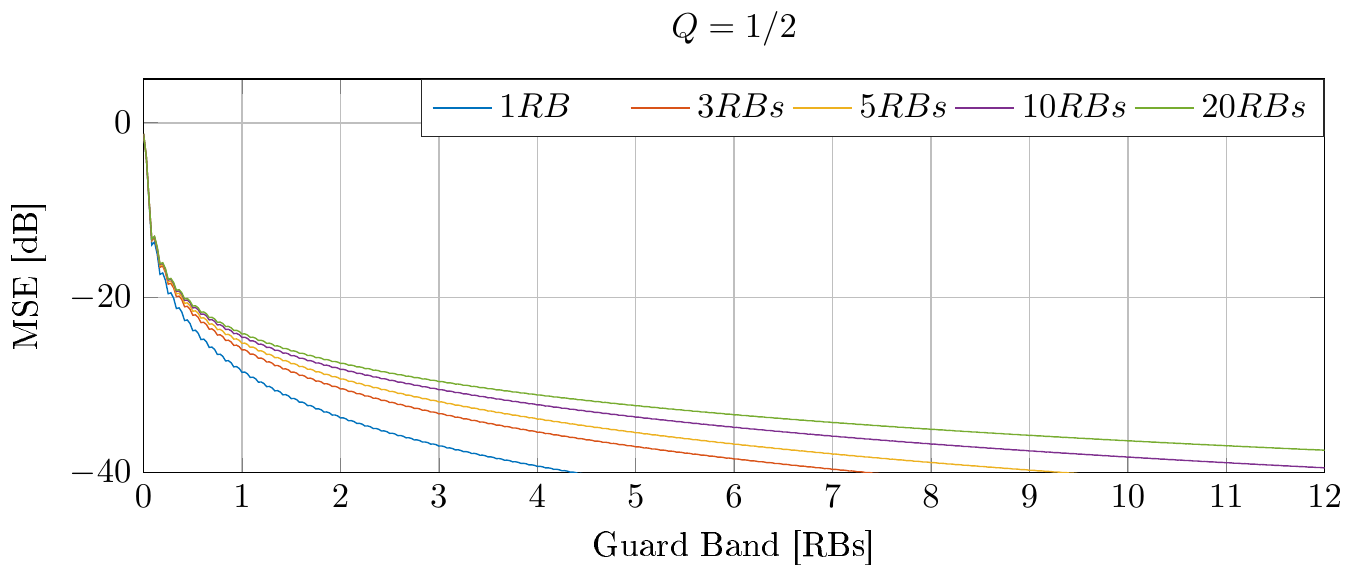}
	\includegraphics[width=0.95\columnwidth, keepaspectratio]{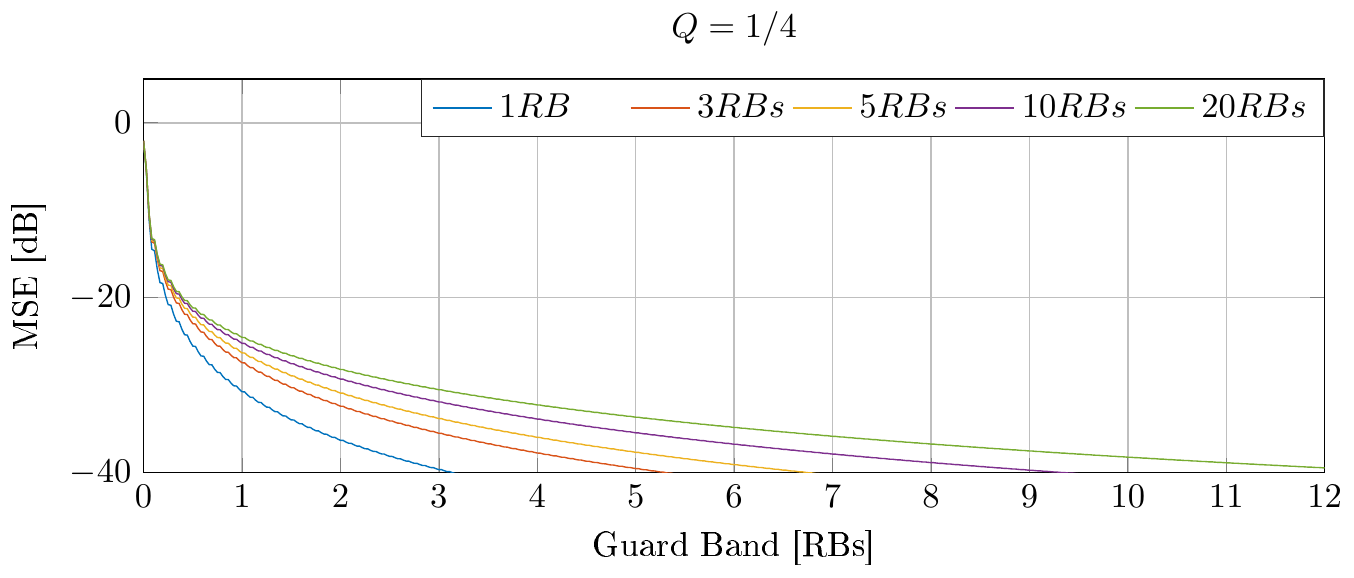}
	\caption{Level of distortion obtained for different interfering bands respectively for Q = \{4,2,1/2,1/4\}.}
	\label{fig:varying-rb}
\end{figure*}

The impact of $N_\mathrm{int}$ on the MSE is now studied. The results are depicted in Figure \ref{fig:varying-rb} for five different interferer bands. 
The obtained results confirm that lower numerologies impact less than higher.  It seems interesting to observe that for a MSE target, the guard band logarithmically increases with the interfering size band.   

\subsection{Guard band dimensioning}
\begin{figure}[!ht]
	\includegraphics[width = 0.95\columnwidth, keepaspectratio]{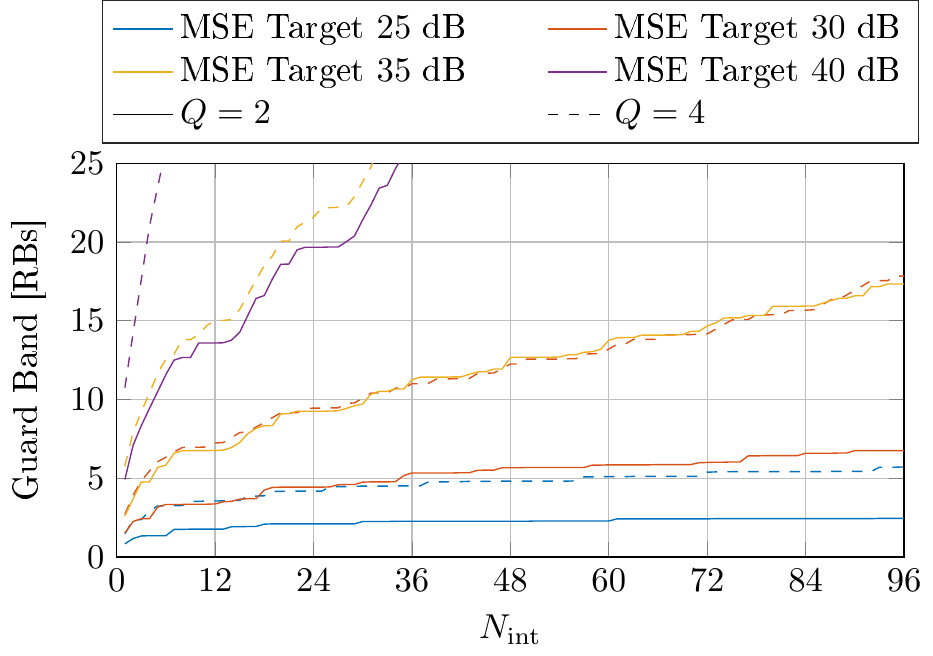}
	\caption{Guard band of function of $N_\mathrm{int}$}
	\label{fig:gb_nint}
\end{figure}

As an application of previous analytical results, it seems interesting to determine the guard band size in order to ensure a given MSE target. The idea is somehow to inverse the expression (\ref{eq:mse_rb_def}) in order to express the guard band as function of the interferer band given a MSE target and a value of $Q$. However, one can observe with Figure \ref{fig:varying-rb} , that the function (\ref{eq:mse-dvlp}) is not bijective. Therefore it can not be inversed. We thus propose to numerically determine the minimum guard band to ensure a given level of distortion for each configuration. The results are depicted in Figure \ref{fig:gb_nint} for $Q= \{2,4\}$. 

Those results confirm the logarithmic trend of the function with respect to the number of interfere subcarriers. As a reminder, the guard band (y axis) is expressed for the user of interest. Therefore the effective guard band [in kHz] is found by multiplying the value by  $180\times 2^{\mu_u}$. Regarding the x axis, the effective interferer band [in kHz] can be found  by multiplying the corresponding value with $15\times 2 ^{\mu_i}$. As expected, the required guard band for $Q=4$ is larger than for $Q=2$. One can observe that the required guard band length levels off quickly for low MSE targets (around $N_{\mathrm{int}} = 50 $ for MSE target = $25$ dB for both $Q=2$ and $Q=4$) but much later for higher MSE targets. It means that a 3-RB-long guard band is enough to ensure $25$ dB MSE for any interferer band with $Q=2$ . However, the required guard band to ensure higher quality becomes significant even for a few RBs of interferer. Indeed, a $9$-RB -long guard band ensures a MSE of  $35$ dB for $2$ RB of interferer with \mbox{$Q =2$} while only $30$ dB MSE for $Q=4$. It means that ensuring low level of interference (high MSE target) is only possible at a significant bandwidth band  efficiency compromise even for $Q=2$. 

\begin{figure}
	\centering
	\includegraphics[width=1\columnwidth, keepaspectratio]{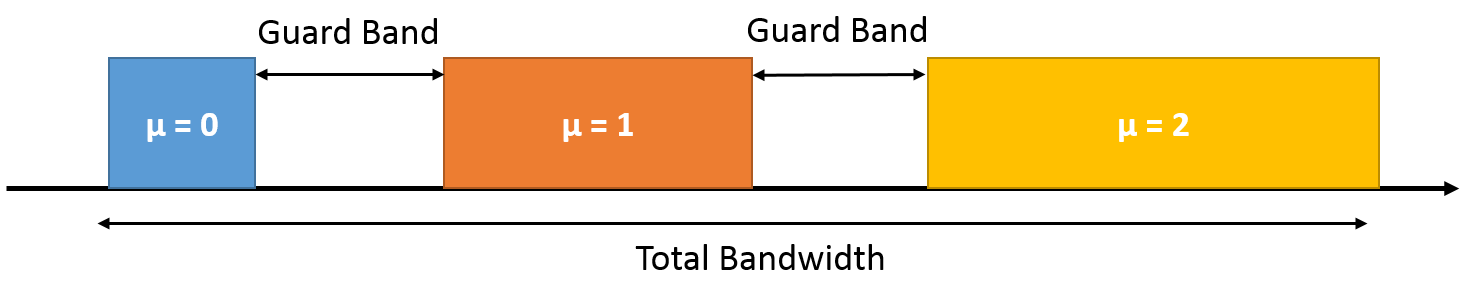}
	\caption{Simple scenario considered for the bandwidth use efficiency evaluation}
	\label{fig:scenario}
\end{figure}

In this last paragraph, we evaluate the bandwidth use efficiency based on a simple scenario as depicted in Figure \ref{fig:scenario}. Three services $\mu=\{0,1,2\}$ are considered and placed as shown such as the numerology ratio between adjacent services is $Q=\{1/2,2\}$. All the services have the same number of allocated RBs. The total bandwidth defined as the sum of the three service bandwidths and the minimum guard band to satisfy the MSE target and the corresponding bandwidth use efficiency are then computed according to the previous results. The results are depicted in Table \ref{tab:band_use}.

One can observe that the bandwidth use efficiency increases with the number of allocated RBs. It is induced by the logarithm trend of the guard band as function of the interferer band observed in Figure \ref{fig:gb_nint}. Besides, the efficiency decreases with the MSE target. It leads to unacceptable bandwidth efficiencies for high quality of service that may be required for ultra-reliable communications. These results justifies the need of side lobe rejection techniques at the transmitter and the receiver side so as to ensure both high inter-numerology interference rejection and acceptable bandwidth use.

\begin{table}[tb]
	\centering
	\caption{Bandwidth use ratio for a given MSE target }
	\label{tab:band_use}
	\begin{tabular}{|l|l|c|c|c|}
		\hline
		\multicolumn{2}{|l|}{\backslashbox{MSE Target}{RBs per service}}                               & 5  RBs & 10 RBs  & 25 RBs\\ \hline
		\multirow{2}{*}{25 dB } & Total Bandwidth {[}MHz{]} & 7.56   & 13.95   & 33.08       \\ \cline{2-5} 
		& Bandwidth use {[}\%{]}      & 83.3   &90.3    & 92.2       \\ \hline \hline
		\multirow{2}{*}{30 dB} & Min Bandwidth {[}MHz{]} & 9.36   &16.29   & 35.78      \\ \cline{2-5} 
		& Bandwidth use {[}\%{]}      &67.3   & 77.35    & 88.05        \\ \hline \hline
		\multirow{2}{*}{40 dB} & Min Bandwidth {[}MHz{]} &  19.85 & 30.65  &  58.23      \\ \cline{2-5} 
		& Bandwidth use {[}\%{]}      &    31.7&    41.1&  54.1    \\ \hline
	\end{tabular}
\end{table}

\section{Conclusion and Perspectives}
The paper investigates the 5G NR eMBB numerology co-existence problem. Based on analytical indicator expressions, general comments and a guard band evaluation have been carried out. Generally speaking, it seems that the key of 5G NR coexistence is to avoid strong multiplexing of the proposed services and to concentrate each service in large dedicated BWPs. However, even by doing so, high inter-numerology interference rejection can only be ensured at the expense of a significant bandwidth use efficiency compromise.  

The proposed interference metric could be part of the future 5G radio resource management algorithm to optimize the user allocation and manage the quality of service given a  level of signal to noise plus interference ratio. 

Eventually, side lobe rejection techniques should therefore be considered in 5G physical layer in order to efficiently support the huge and diverse traffic. Recent literature has focused on filtering and/or windowing to deal with asynchronous communications. Those solutions may be applied and extended to the 5G NR numerology coexistence scenario. 

%

\bibliographystyle{IEEEtran}
\bibliography{bibliography}




\end{document}